\documentclass[10pt]{article} 
\usepackage[english]{babel}
\usepackage[utf8]{inputenc}
\usepackage{geometry}
\usepackage{mathtools}
\usepackage{amsmath}
\usepackage{amsfonts}
\usepackage{hyperref}
\usepackage[ruled,vlined]{algorithm2e}
\usepackage{caption}
\usepackage{subcaption}
\usepackage{bm}
\usepackage{bbm}
\usepackage{amssymb}
\newtheorem{definition}{Definition}
\usepackage{graphicx}
\usepackage{setspace}
\usepackage{adjustbox}
\usepackage[backend=bibtex, style=authoryear, maxcitenames=2, maxbibnames=99, dashed=false]{biblatex}

\usepackage [english]{babel}

\begin{document}

\title{Designing Proposal Distributions for Particle Filters using Integrated Nested Laplace Approximation}

\author{Alaa Amri$^{a}$ \footnote{Corresponding author: Alaa, Amri; \tt{alaa.amri@ed.ac.uk}}.\\
        \small $^{a}$School of Mathematics, The University of Edinburgh (United Kingdom)\\
}

\date{} 

\maketitle

\begin{abstract} 
\noindent State-space models are used to describe and analyse dynamical systems. They are ubiquitously used in many scientific fields such as signal processing, finance and ecology to name a few. Particle filters are popular inferential methods used for state-space methods. Integrated Nested Laplace Approximation (INLA), an approximate Bayesian inference  method, can also be used for this kind of models in case the transition distribution is Gaussian. We present a way to use this framework in order to approximate the particle filter’s proposal distribution that incorporates information about the observations, parameters and the previous latent variables. Further, we demonstrate the performance of this proposal on data simulated from a Poisson state-space model used for count data. We also show how INLA can be used to estimate the parameters of certain state-space models (a task that is often challenging) that would be used for Sequential Monte Carlo algorithms. \end{abstract}

\begin{flushleft}
\end{flushleft}

\section{Introduction}

Particle filters are sequential Monte Carlo methods used for Bayesian inference in state-space models. It is crucial to carefully design the \emph{proposal distribution} of the particle filter: the distribution from which particles or samples are drawn, as it significantly affects the performance of the particle filter. In particular, the design affects the variance of marginal likelihood estimates and attempts to address the issue of \emph{particle degeneracy}, a phenomenon where most samples have extremely low normalized importance weights and eventually do not contribute to the approximation of the target distribution. This challenge becomes particularly important in high-dimensional state-space models.

\subsection{State of the art}
The bootstrap particle filter \parencite{bpf}, a widely used variant, draws particles directly from the prior distribution or transition distribution. However, a notable limitation of this approach is that particles drawn at time \( t \), denoted \( h_{t} \), are not necessarily aligned with the corresponding observed data point \( y_{t} \). As a result, this can lead to inefficiencies in representing the target distribution accurately. 

To address these challenges, several extensions to the bootstrap particle filter algorithm have been developed. One example is the Rao-Blackwellised Particle Filter (RBPF) \parencite{rbpf}, which improves estimation accuracy by marginalizing out specific components of the state process, applying the particle filter only to a lower-dimensional component. This approach reduces variability in the estimates but is primarily applicable to models with a partially linear and Gaussian structure \parencite{johansen2012exact}.  There is also the auxiliary particle filter that was introduced in \textcite{apf} to improve upon the bootstrap particle filter especially when the data points are very informative i.e the signal-to-noise ratio is relatively high and it includes an auxiliary variable construction. Various improvements to the auxiliary particle filter are available like \parencite{apf2,apf5,apf3,apf4,johansen2008note,whiteley2010recent}.

The \emph{optimal} proposal distribution $p(h_t \mid h_{t-1}, y_t) $ minimizes the  variance of the importance weights conditional on all previous latent variables $h_{1:t-1}$ and all observations up to time step $t$, $y_{1:t}$ \parencite{int3}. Apart from a few models such as the linear Gaussian state-space model, sampling from the optimal distribution is not possible. Hence, Laplace approximation can be used to locate the modes of the unnormalized optimal sampling distribution, assuming it is log-concave, so that a Gaussian or an over-dispersed $t$-distribution is fitted around each of these modes. Locating the modes requires using an optimisation algorithm, such as the Newton-Raphson method, for each particle which is not practical from a computational perspective \parencite{int0} and \parencite[Chapter 4]{int1}. Further, making a Gaussian approximation can also be done using a Taylor expansion but it is not efficient if the optimal kernel is heavy-tailed or not close to a Gaussian \parencite[Chapter 10]{Chopin2020}.

Another possible way to approximate the (optimal) proposal distribution is to use the extended Kalman filter (EKF) or the unscented Kalman filter (UKF) \parencite{int5,int4} where the latter yields more accurate and stable estimates when compared to the former. In the case of the extended Kalman filter, the linearisation of the observation and state dynamics, is carried out, using Jacobian matrices that are computed and evaluated at each time step whereas the unscented Kalman filter uses a deterministic sampling of carefully selected points (sigma points) to make the approximation. The unscented Kalman filter yields more accurate and stable estimates when compared to the extended Kalman filter.  In \textcite{defreitas2000sequential}, the proposal distribution for the particle filter (PF) is based on the EKF Gaussian approximation. In \textcite{vandermerwe2000unscented}, the authors adopt a similar approach but substitute the EKF proposal with a UKF proposal. In both cases, the process of linearisation is performed for every
individual particle. 

\subsection{Proposed method and contributions}

In this paper, we focus on approximating $p_{\theta}\left(h_t \mid h_{1:t-1}, y_{1: T} \right)$ (and $p_{\theta}\left(h_1 \mid  y_{1: T}\right)$ in case of $t=1$), instead of approximating the optimal proposal distribution, to use as the proposal in the particle filter. To accomplish this, we employed integrated nested Laplace approximations (INLA) \parencite{inla1} for the approximation. INLA has gained popularity as an inference method in recent years, especially in fitting spatial and/or spatio-temporal models. It uses a combination of analytical approximations and efficient numerical integration schemes to achieve highly accurate deterministic approximations to posterior distributions. This method enables fast inference performance in latent Gaussian Models. Moreover, it is considered as a well-known alternative to sampling-based methods such as Markov chain Monte Carlo (MCMC) algorithms to perform Bayesian inference. The main two steps consist of constructing a Gaussian approximation to the joint posterior of the parameters conditional on the data then approximating the marginals of the latent variables given the data and the parameters. For more details, see \parencite{inla01,inla02,inla03} and references therein. Some of the uses of INLA for state-space models, include \parencite{inla04,inla05}.

Unlike the previously mentioned studies that relied on the Laplace approximation, our approach does not require performing an approximation for each individual particle in the SMC process, which results in significant computational speedup. Additionally, INLA offers a more refined and versatile solution compared to the Laplace approximation, as it incorporates more advanced numerical integration techniques, providing improved accuracy and efficiency in complex models. Moreover, INLA exploits the sparsity structure to accelerate computations. This work represents the first attempt to combine integrated nested Laplace approximations (INLA) with sequential Monte Carlo (SMC) methods. The other novelty consists of designing a proposal distribution based on approximations of marginal smoothing distributions $p_{\theta}\left(h_t \mid  y_{1: T}\right)$ and the complete smoothing distribution $p_{\theta}\left(h_{1: T} \mid y_{1: T}\right)$ as in our case.

\subsection{Outline of the paper}
The remainder of this paper is organised as follows. 
Section \ref{preliminaries} presents a formal definition of state-space models and it recalls how particle filters (PFs) and Integrated nested Laplace approximation (INLA) work. 
In Section \ref{inlachapsec2}, we present our contribution which consists of designing the proposal distributions of the particle filter from the approximated probability distributions obtained by the INLA method. Section \ref{inlachapsec3} contains experiments that illustrate the performance of the proposed method, based on simulated data from a Poisson state-space model. Additionally, it includes another example that shows how INLA can also be used for parameter estimation of state-space models or can be combined with one of the methods that is often used for this task, namely particle MCMC. The paper concludes with remarks in Section \ref{conclusions}.

\section{Preliminaries}
\label{preliminaries}
\subsection{State-space models}
\begin{flushleft}
    For a given time horizon $T$ (total number of observations), we have $\mathcal{X}$-valued $\{ H_t \}_{t=1}^{T}$ 
and $\mathcal{Y}$-valued $\{ Y_t \}_{t=1}^{T}$ two stochastic processes representing the hidden variables and observations respectively. A state space model can be expressed as follows
\end{flushleft}

\begin{equation} 
\label{ssmodel}
 \begin{split}
     Y_{t}|H_{t} &= h_{t}   \sim p_\theta(y_t \mid h_t), \\
     H_{t} \mid H_{t-1} &= h_{t-1}  \sim  p_\theta(h_t \mid h_{t-1}),\\
      H_{1} &= h_{1} \sim p_\theta(h_1), 
 \end{split}
\end{equation}

\begin{flushleft}
     where $\theta \in \Theta$  is the set of the model’s
parameters, $p_\theta(h_1)$ is the initial distribution defining the prior distribution of the hidden state $H_1$ at the starting time \( t = 1 \), $p_\theta(h_t \mid h_{t-1})$ is called the transition distribution and it describes how latent variables evolve from the time step $t-1$ to $t$ in case of not observing any data and $p_\theta(y_t \mid h_t)$ provides the probabilities of observing $Y_t$ conditionally on $H_{t} = h_{t} $. 
\end{flushleft}

\subsection{Particle filters}
 At each time step \( t \), particle filters involve sampling particles and computing their associated weights $w_{t}$, which are then normalized and stored to estimate desired quantities or approximate posterior distributions, such as filtering distributions, the normalised weights are denoted by $W_{t}$. To improve efficiency, the resampling step is performed adaptively, specifically when the weight variability becomes excessive. This approach, known as adaptive resampling, is triggered when the Effective Sample Size (ESS) drops below a predefined threshold, \( ESS_{min} \). Algorithm \ref{alg1.2} is a pseudo-code of a generic particle filter. It is important to highlight that operations with the superscript $i$ are carried out for each individual particle among the N particles.

\begin{algorithm}[!ht]
\DontPrintSemicolon
\tcp{Operations involving the superscript $i$ are performed for $i = 1,\dots,N$.}

Sample $H_1^{i} \sim q_{\theta}(h_1)$\;

Compute weights \[
w_{1}^i = \frac{p_\theta\left(h_{1}=H_{1}^{i}\right) p_\theta\left(y_1 \mid h_{1}=H_{1}^{i}\right)}{q_{\theta}\left(h_{1}=H_{1}^{i}\right)}.
\]

Normalize weights \[
W_{1}^i = \frac{w_{1}^i}{\sum_{j=1}^{N} w_{1}^j}.
\]

\For{$t = 2,\dots,T$}{
    Calculate $\text{ESS} := \frac{1}{\sum_{i=1}^{N} \left(W_{t-1}^{i}\right)^{2}}$\;
    
    \If{$\text{ESS} \leq \text{ESS}_{\text{min}}$}{
        Draw the index $a_{t}^{i}$ of the ancestor of the particle $i$, by resampling the normalized weights $W_{t-1}^{1:N}$\;
        Set $\hat{w}^{i}_{t-1} = 1$\;
    }
    \Else{
        $a_{t}^{i} = i$\;
        $\hat{w}^{i}_{t-1} = w_{t-1}^{i}$\;
    }

    Sample $H_t^{i} \sim q_{\theta}(h_t \mid h_{t-1}=H_{t-1}^{a_{t}^{i}})$\;

    Compute weights \[
    w_{t}^i = \frac{p_\theta\left(h_{t}=H_{t}^{i} \mid h_{t-1}=H_{t-1}^{a_{t}^{i}}\right) p_\theta\left(y_t \mid h_{t-1}=H_{t-1}^{a_{t}^{i}}\right)}{q_{\theta}\left(h_{t}=H_{t}^{i} \mid h_{t-1}=H_{t-1}^{a_{t}^{i}}\right)} \hat{w}^{i}_{t-1}.
    \]

    Normalize weights \[
    W_{t}^i = \frac{w_{t}^i}{\sum_{j=1}^{N} w_{t}^j}.
    \]
}
\caption{Generic particle filter}
\label{alg1.2}
\end{algorithm}

\begin{flushleft}
    In this work, we used systematic resampling. Empirically, it performs better than other resampling schemes such as stratified resampling and especially multinomial resampling, as it yields estimates with lower variability \parencite[Chapter 9]{Chopin2020}. It is worth noting that a by-product of the particle filter’s output is an unbiased estimate of the marginal likelihood $p_{\theta^{}}\left(y_{1: T}\right)$   \parencite{delmoral}.  It was shown by \textcite{delmoral2} that the estimate $\hat{p}_{\theta^{}}\left(y_{1: T}\right)$ has a non-asymptotic variance that grows linearly with the time horizon T. 
\end{flushleft}
 
\subsection{Integrated Nested Laplace Approximation}
\label{inlachapsec1}
\begin{flushleft}
    In this part, we provide a description of the integrated nested Laplace approximation (INLA) methodology which is used for latent Gaussian models (LGMs), a model formulation
that consists of a hierarchical model with three levels or layers. Let us consider the case where we have observations $y = (y_1,y_2,\ldots,y_T) \in \mathbb{R}^{T\times d}$ with corresponding random variables $Y = (Y_1,Y_2,\ldots,Y_T)$. The latent Gaussian model is the hierarchical Bayesian model where we have
\end{flushleft}

\begin{equation} 
Y \mid H, \theta \sim \prod_{i=1}^T p\left(y_i \mid h_i, \theta\right), \\
 \end{equation}

\begin{flushleft}
    where $\theta$ is the $d_{\theta}$-variate vector of hyperparameters, and $H$ represents the latent variables and also known as \textit{the latent field}. It is assumed that $Y$ are conditionally independent given $H$ and $\theta$. INLA \parencite{inla1} can be used for latent Gaussian models by assuming that the latent field $H$ is Gaussian Markov Random field \parencite{inla2} with mean $\mathbf{0}$, a $(T\times d)$-length vector of zeros, and precision matrix $\textbf{Q}$. 
\end{flushleft}

\begin{definition}
\label{gmrf}
    \parencite{inla2} Let $H=\left(H_1, \ldots, H_T\right)^{'} \in \mathbb{R}^T$ have a normal distribution with mean $\mu$ and precision matrix $\textbf{Q}$. Define the labelled graph $\mathcal{G}=(\mathcal{V}, \mathcal{E})$, where vertices $\mathcal{V}=$ $\{1, \ldots, T\}$ and edges $\mathcal{E}$ be such that there is no edge between node $i$ and $j$ iff $H_i \perp H_j \mid H_{-i j}$. $H$ is called a Gaussian Markov Random Field (GMRF) w.r.t a labelled graph $\mathcal{G}=(\mathcal{V}, \mathcal{E})$ with mean $\mu$ and precision matrix $\textbf{Q}$, iff its density has the following form:
$$
(2 \pi)^{-T / 2}|\textbf{Q}|^{1 / 2} \exp \left(-\frac{1}{2}(h-\mu)^{'} \textbf{Q} (h-\mu)\right)
$$
and
$$
\textbf{Q}_{i j} \neq 0 \quad \Longleftrightarrow \quad\{i, j\} \in \mathcal{E} \quad \text { for all } i \neq j. 
$$ 
\end{definition}

\begin{flushleft}
    Hence, the latent field $H$ and the parameters $\theta$  are given below
\end{flushleft}
\begin{equation} \begin{aligned}
    \begin{aligned}
H \mid \theta & \sim \mathcal{N}\left(\mathbf{0}, \textbf{Q}^{-1}\right), \\
\theta & \sim p(\theta).
\end{aligned}
\end{aligned} \end{equation}




INLA seeks to create numerical approximations to the marginal posterior distribution of the latent field, $p\left(h_i \mid y\right)$ and the hyperparameters, $p\left(\theta_j\mid y\right)$ instead of approximating the joint posterior distribution $p(h, \theta \mid y)$. Operationally, INLA proceeds in the following steps

\paragraph{Step 1:}
  Approximate $p\left(\theta \mid y\right)$ using Laplace approximation. 
  
The posterior of the hyperparameters  given the observations, 
$p(\theta \mid y)$ is approximated by $p_{LA}(\theta \mid y)$, as follows:

\begin{equation} \begin{aligned}
p_{}(\theta \mid y) & =\frac{p(h, \theta \mid y)}{p(h \mid \theta, y)} \propto \frac{p(\theta) p( h \mid \theta) p(y \mid h, \theta)}{p(h \mid \theta, y)} \\
 &\left.\approx \frac{p(\theta) p( h \mid \theta) p(y \mid h, \theta)}{p_{G}(h \mid \theta, y)} \right|_{h = m^{*}}
 :=p_{LA}(\theta \mid y),
\end{aligned} 
\label{inlaapproxeq1}
\end{equation}

where the Gaussian approximation \( p_G(h \mid \theta, y) \) of \( p(h \mid \theta, y) \) is derived by matching the mode and curvature around the mode \( m^* \). The mode of \( p_{LA}(\theta \mid y) \) is found using an iterative optimization algorithm like Newton-Raphson, followed by exploration of the surrounding area. A grid search identifies integration points \( \{\theta^{(s)}\}_{s=1}^S \) and corresponding weights \( \{\Delta_s\}_{s=1}^S \), enabling numerical integration to approximate the distribution. Finally, marginalizing \( p_{LA}(\theta \mid y) \) provides the marginal distributions for each parameter \( p_{LA}(\theta_i \mid y) \).

\paragraph{Step 2:} Approximate $p\left(h_i \mid \theta, y\right)$.

\begin{equation} \begin{aligned}
\left. p\left(h_i \mid \theta, y\right) \approx \frac{p\left(h, \theta \mid y\right)}{p_{G}\left(h_{-i} \mid h_i, \theta, y\right)}\right|_{h_{-i}=m_{-i}} := p_{LA}\left(h_i \mid \theta , y\right),
\end{aligned} 
\label{inlaapproxeq2}
\end{equation}

\begin{flushleft}
where $p_{G}\left(h_{-i} \mid h_i, \theta, y\right)$ is the Gaussian approximation of $p_{}\left(h_{-i} \mid h_i, \theta, y\right)$ and $m_{-i}$ is its mode. \textcite{inla1} introduced a faster alternative called the simplified Laplace approximation. This method approximates \(\log p(h_i | \theta, y)\) using a Taylor series expansion around the mode, including up to third-order terms. To further improve accuracy, a multivariate skew-normal distribution is fitted to account for errors in the location and skewness. Consequently, \(p(h_i | \theta, y)\) is approximated as a mixture of skew-normal distributions \parencite{azzalini1999statistical,rue2017bayesian}. 
\end{flushleft}

\paragraph{Step 3:} Approximate $p\left(h_i \mid  y\right)$.

\begin{equation} \begin{aligned}
p\left(h_i \mid y\right)&=\int p\left(h_i \mid \theta, y\right) p(\theta\mid y) \mathrm{d} \theta.  \\
&\approx \sum_{s=1}^S p_{LA}\left(h_i \mid \theta=\theta^{(s)} , y\right) p_{LA}\left(\theta=\theta^{(s)} \mid y\right) \Delta_s.
\end{aligned} \end{equation}

\begin{flushleft}
    The approximation of \( p(h_i \mid y) \) combines the two previous approximations (\eqref{inlaapproxeq1} and \eqref{inlaapproxeq2}) and integrates out \( \theta \) numerically. This is done using the integration points \( \{\theta^{(s)}\}_{s=1}^S \) and their associated weights \( \{\Delta_s\}_{s=1}^S \).
\end{flushleft}

\section{Proposed method: using INLA Approximations for Proposal Distributions in Particle Filters}
\label{inlachapsec2}

\begin{flushleft}

We now describe how to leverage INLA to propose a novel particle filter based on the approximated distributions using the INLA method in the case where the state-space model can be expressed as a latent Gaussian model. Algorithm \ref{alg2.1} provides a thorough outline of the procedure.

We require INLA to solely obtain a Gaussian approximation of $p_{\theta}\left(h_{1: T} \mid y_{1: T}\right)$ as in \eqref{inlaapproxeq1} and an approximation to $p_\theta\left(h_{1} \mid y_{1: T}\right)$. Let us assume that $h_{t} \in \mathbb{R}^d$ and $y_{t} \in \mathbb{R}^d$ for all $t = 1,..,T$.
Here we have:    
\end{flushleft}

\begin{equation} \begin{aligned}
\label{eq2.4}
p_{\theta}\left(h_{1: T} \mid y_{1: T}\right) \approx \mathcal{N} \left(\mu_{1:T} , \Sigma_{1:T} \right),
\end{aligned} \end{equation}

\begin{flushleft}
 where $\mu_{t} \in \mathbb{R}^d$ for all $t = 1,..,T$, $\mu_{1:T} = (\mu_1,..,\mu_T)^{'} \in \mathbb{R}^{Td}$, $\Sigma_{t}$ (for all $t \leq T$) and $\Sigma_{1:T}$ are respectively  $d \times d$ and $Td \times Td$ covariance matrices that are symmetric and positive definite. Note that  (\ref{eq2.4}) can also be expressed as follows:
\end{flushleft}

\begin{equation} \begin{aligned}
\begin{split}
p_{\theta}\left(h_{1: T} \mid y_{1: T}\right) &=  p_{\theta}\left(h_{1: T-1}, h_T \mid y_{1: T}\right) \\
&= p_{\theta}\left(h_T \mid h_{1: T-1}, y_{1: T}\right) p_{\theta}\left(h_{1: T-1} \mid y_{1: T}\right)
\\
&\approx \mathcal{N}\left(\left(\begin{array}{l}
\mu_{1:T-1} \\ 
\mu_T
\end{array}\right),\left(\begin{array}{cc}
\Sigma_{1:T-1} & C_{1:T-1 \mid T} \\
C^{\top}_{1:T-1 \mid T} & \Sigma_{T}
\end{array}\right)\right),
\end{split}
\end{aligned} \end{equation}

\begin{flushleft}
    where $\mu_{1:T}$ and $\Sigma_{1:T}$ are subdivided into blocks. Here, $\mu_{1:T-1} = (\mu_1,..,\mu_{T-1})^{'}$, $\Sigma_{1:T-1}$ and $C_{1:T-1 \mid T}$ are $(T-1)d \times (T-1)d$ and $(T-1)d \times d$ matrices respectively.
    
    We first demonstrate the case of the last time step T where we approximate   $p_{\theta}\left(h_T \mid h_{1:T-1}, y_{1: T} \right)$ as follows:
\end{flushleft}

\begin{equation} \begin{aligned}
\label{eq4.1}
p_{\theta}\left(h_T \mid h_{1: T-1}, y_{1: T} \right) \approx \mathcal{N} \left(\Tilde{\mu}_T, \Tilde{\Sigma}_{T} \right),
\end{aligned} \end{equation}
where computing the mean $\Tilde{\mu}_T$ and the variance $\Tilde{\Sigma}_{T}$ 
proceeds by performing the following calculations:
\begin{equation} \begin{aligned}
\label{eq4.3}
\begin{split}
\Tilde{\mu}_T &= \mu_T +  C^{\top}_{1:T-1 \mid T} \Sigma^{-1}_{1:T-1} (h_{1: T-1}-\mu_{1:T-1}),\\
 \Tilde{\Sigma}_{T} &= \Sigma_{T} - C^{\top}_{1:T-1 \mid T} \Sigma^{-1}_{1:T-1}C_{1:T-1 \mid T}.
 \end{split}
\end{aligned} \end{equation}

\begin{flushleft}
      Note that an approximation of $p_{\theta}\left(h_t \mid h_{1:t-1}, y_{1: T} \right)$ can be obtained, recursively, in a similar way as in (\ref{eq4.1})-(\ref{eq4.3}), for all $t = 2,..,T$. As for $t = 1$, we consider the approximation of the marginal distribution $p_\theta\left(h_{1} \mid y_{1: T}\right)$ 
 obtained directly from INLA using \eqref{inlaapproxeq2}. We propose using these approximated distributions as proposal distributions for the particle filter. Unlike the transition distributions used in the bootstrap particle filter, these distributions account for all available observations, making them different from the standard approach. At each time step \( t \), each $i$-th particle is assigned a weight $w_{t}^i$, which is then normalized to produce $W_{t}^i$, the normalized weight. If necessary, these normalized weights $W_{t}^{1:N}$ are used for resampling the particles.
\end{flushleft}

\begin{algorithm}[!ht]

\DontPrintSemicolon

   \tcp{Operations involving the superscript i are performed
for $i =$ 1,...,N.}
   
  Fit the model using INLA and store $\mu_{1:T}$ and  $\Sigma_{1:T}$.
  
  Sample $H_1^{i} \sim q_{\theta}(h_1) \approx$  $p_\theta\left(h_{1} \mid y_{1: T}\right)$.\\

  Compute weights \qquad $w_{1}^i$  $= \frac{p_\theta\left(h_{1}=H_{1}^{i}\right) p_\theta\left(y_1 \mid h_{1}=H_{1}^{i}\right)}{q_{\theta} \left(h_{1}=H_{1}^{i}\right)}$.\\

  Normalize weights \qquad $W_{1}^i=$  $\frac{w_{1}^i}{\sum_{j=1}^{N} w_{1}^j}$.

  \For{t = 2,..,T}{

   Draw the index $a_{t}^{i}$ of the ancestor of the particle i, by resampling the normalised weights $W_{t-1}^{1:N}$.\\

   Compute $\Tilde{\mu}_t^i = \mu_t +  C^{\top}_{1:t-1 \mid t} \Sigma^{-1}_{1:t-1} (H_{1: t-1}^{a_{t}^{i}}-\mu_{1:t-1})$.\\
 Compute $\Tilde{\Sigma}_{t}= \Sigma_{t} - C^{\top}_{1:t-1 \mid t} \Sigma^{-1}_{1:t-1}C_{1:t-1 \mid t}.$ \\
   
   Sample $H_t^{i} \sim \mathcal{N} \left(\Tilde{\mu}_t^i , \Tilde{\Sigma}_{t} \right)$.\\

   Compute weights \qquad $w_{t}^i$ $= \frac{p_\theta\left(h_t= H_t^i|h_{t-1}=H_{t-1}^{a_{t}^{i}}\right) p_\theta\left(y_t \mid h_t= H_t^i\right)}{q_{\theta} \left(h_t= H_t^i|h_{t-1}=H_{t-1}^{a_{t}^{i}}\right)}$.\\

   Normalize weights \qquad $W_{t}^i=$ $\frac{w_{t}^i}{\sum_{j=1}^{N} w_{t}^j}$.

  }

\caption{Particle Filter with INLA-approximated proposals}
\label{alg2.1}
\end{algorithm}

\subsection{Discussion}
It is important to note that Algorithm \ref{alg2.1} can be integrated into the particle marginal Metropolis-Hastings (PMMH) sampler \parencite[]{pmcmc}. The resulting procedure starts by initialising parameters using $\theta^{(0)}$ and running a INLA-based particle filter \ref{alg2.1} to estimate the marginal likelihood and obtain a sample of the latent variables $H_{1: T}^{(0)}$. At each iteration $k$, a new parameter $\theta^{'}$ is proposed from the proposal distribution $q \left( \theta \mid \theta= \theta^{(k-1)}\right)$, and another INLA-based particle filter is employed based on $\theta^{'}$ to obtain a new estimate of the marginal likelihood $p_{\theta^{'}}\left(h_{1: T} \mid y_{1: T}\right)$ and another sample of the latent variables $\hat{p}_{\theta^{'}}\left(y_{1: T}\right)$. The acceptance probability is then computed by comparing the likelihoods’ estimates, parameters’ priors and proposal densities of the current and proposed parameter values. In case of acceptance, both the parameters $\theta^{'}$  and the latent states $H_{1: T}^{'}$ are stored; otherwise, the parameters $\theta^{(k-1)}$ and the corresponding latent states $H_{1: T}^{(k-1)}$from the previous iteration are retained. The resulting procedure is outlined in Algorithm \ref{pmmh_alg2}. While INLA can serve as a method for parameter estimation in state-space models, we can also initialize the Markov chains of the PMMH samplers using the modes from \eqref{inlaapproxeq1}. \\

To further decrease computational time, the simplified Laplace approximation can be employed in place of the full Laplace approximation when working with approximation $p_\theta\left(h_{1} \mid y_{1: T}\right)$. On a related note, $p_\theta\left(h_{1} \mid y_{1: T}\right)$ is also referred to as the marginal smoothing distribution at time 1, which can be approximated using particle smoothing algorithms. When compared to particle filtering, particle smoothing is generally considered more challenging. While some of these challenges are computational, there are others that are related to how restrictive some algorithms are (they are solely applied to a specific class of state-space models).
\parencite[Chapter 12]{Chopin2020} provides a thorough treatment on a plethora of these algorithms in addition to highlighting their characteristics such as the computational complexity per time step.

\begin{algorithm}[h]
\DontPrintSemicolon

\textbf{Input:} $K, N, \theta^{(0)}$\\

Run Algorithm \ref{alg2.1} targeting $p_{\theta^{(0)}}\left(h_{1: T} \mid y_{1: T}\right)$, store a sample $H_{1: T}^{(0)}$ and let $\hat{p}_{\theta^{(0)}}\left(y_{1: T}\right)$ denote the marginal likelihood estimate. \\
  \For{k = 1,..,K}{

Sample $\theta^{'} \sim$  $q \left( \theta \mid \theta= \theta^{(k-1)}\right)$,\\
Run Algorithm \ref{alg2.1}  targeting $p_{\theta^{'}}\left(h_{1: T} \mid y_{1: T}\right)$, store a sample $H_{1: T}^{'}$ and let $\hat{p}_{\theta^{'}}\left(y_{1: T}\right)$ denote the marginal likelihood estimate.\\
With probability,
  \\
$$
min \left( 1 , \frac{\hat{p}_{\theta^{'}} \left(y_{1: T}\right) p\left(\theta = \theta^{'}\right) q\left(\theta^{(k-1)} \mid \theta^{'}\right)}{\hat{p}_{\theta^{(k-1)}}\left(y_{1: T}\right) p\left(\theta = \theta^{(k-1)}\right) q\left(\theta^{'} \mid \theta^{(k-1)}\right)}\right).
$$ 
  \\
Set $\theta^{(k)}=\theta^{'}$, $H_{1: T}^{k}=H_{1: T}^{'}$ and $\hat{p}_{\theta^{(k)}}\left(y_{1: T}\right)=\hat{p}_{\theta^{'}}\left(y_{1: T}\right)$; otherwise set $\theta^{(k)}=\theta^{(k-1)}$, $H_{1: T}^{k}=H_{1: T}^{k-1}$ and $\hat{p}_{\theta^{(k)}}\left(y_{1: T}\right)=\hat{p}_{\theta^{(k-1)}}\left(y_{1: T}\right)$.
}

\caption{Particle Marginal Metropolis–Hastings with INLA-based particle filter}
\label{pmmh_alg2}
\end{algorithm}

\section{Numerical experiments}
\label{inlachapsec3}

\begin{flushleft}
In this section, we present empirical evidence of the effectiveness of the suggested approach by comparing it with the bootstrap particle filter for simulated data. 
    We considered the following state-space model:
\end{flushleft}
\begin{equation} \begin{aligned}
\label{model}
 \begin{split}
     Y_{t}|H_{t} &= h_{t}   \sim Poisson(e^{h_{t}+\alpha}), \\
     H_{t} \mid H_{t-1} &= h_{t-1}  \sim \mathcal{N}\left(\rho h_{t-1}, \sigma^{2} \right), \\
      H_{1} &= h_{1} \sim \mathcal{N}\left(0, \frac{\sigma^{2}}{1- \rho^{2}} \right),
 \end{split}
\end{aligned} \end{equation}

\begin{figure}[h]

         \centering
         \includegraphics[width=0.8\linewidth]{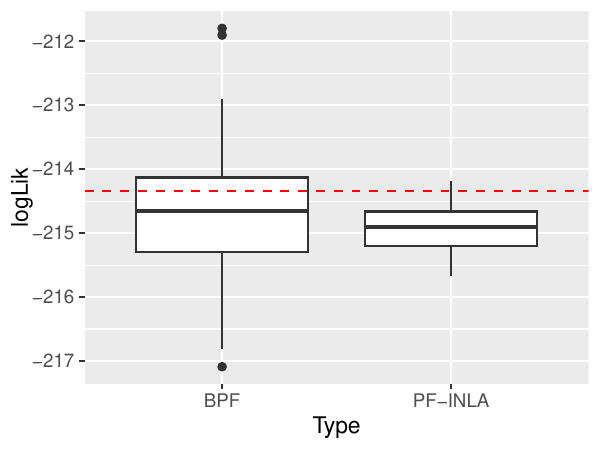}

	    	\caption{ Box plots of estimates of log likelihood, log $p_\theta\left(y_{1: T}\right)$, obtained by the two particle filters: bootstrap (left) and INLA-based (right) (over 50 runs for $N=100$ and $T=100$). The red dotted horizontal line denotes the log of the marginal likelihood estimate obtained from a bootstrap particle filter with $N=10^{6}$.}
      \label{fig:3.01}
\end{figure}

\begin{figure}[h]
 
         \centering
         \includegraphics[height = 6cm,width=1\linewidth]{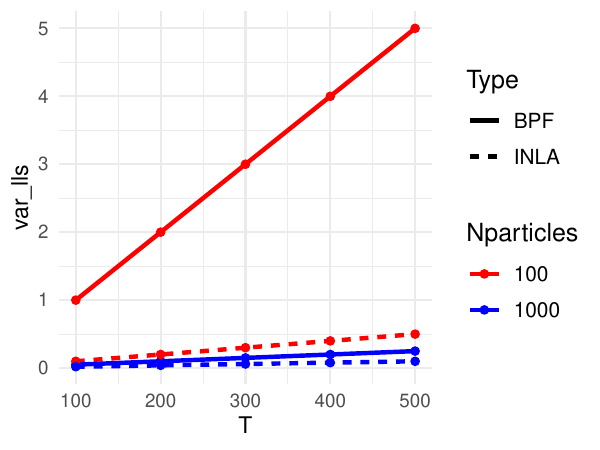}

	    	\caption{The variance of the log likelihood estimates (over 50 runs) by the two particle filters (dotted lines for the INLA-based and straight lines for the bootstrap), as a function of the number of particles $N$ (for both $N = 100$ (in red) and $N = 1000$ (in light blue) and the time horizon $T$ (from $T=100$ to $T=500$).}
      \label{fig:3.002}
\end{figure}

\begin{figure}[]
 \begin{subfigure}[]{\textwidth}
         \centering
         \includegraphics[width=1\linewidth]{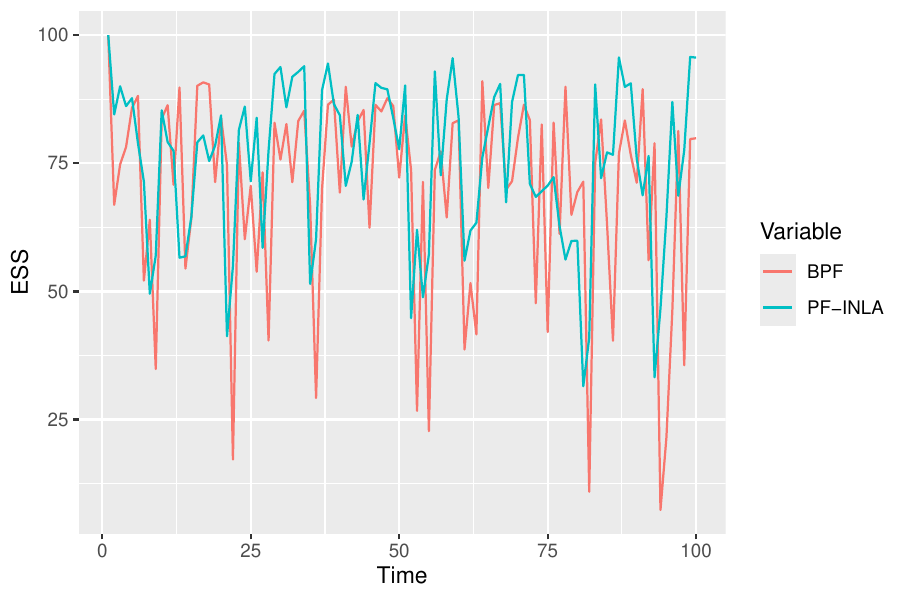}
         \caption{}
     \end{subfigure}
      \begin{subfigure}[]{\textwidth}
         \centering
         \includegraphics[width=1\linewidth]{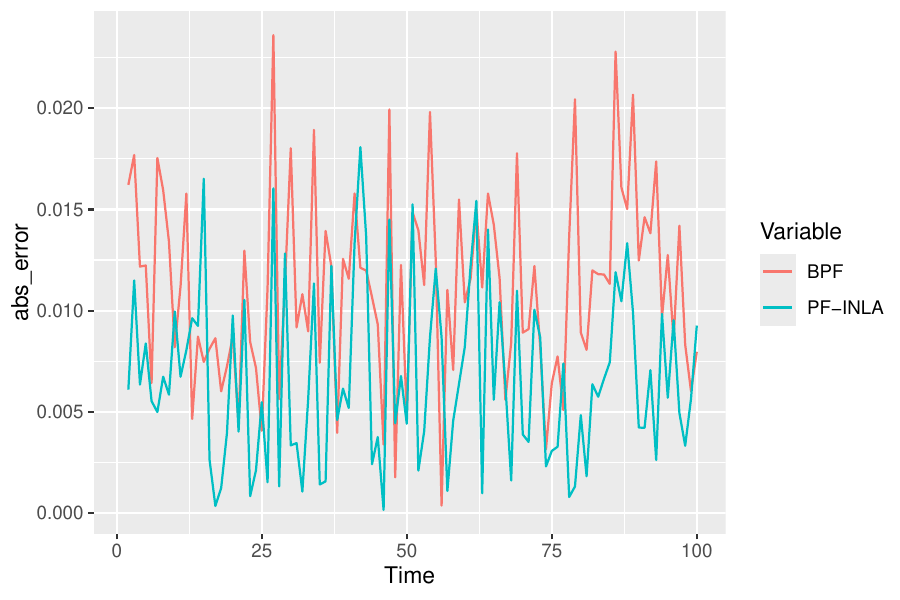}
         \caption{}
     \end{subfigure}

	    	\caption{a) Average of the Effective Sample Size (ESS) as a function of time t (over 50 runs) of the two particle filters ($N=100$ and $T=100$). \\ b) The average of particle estimate absolute errors as a function of t for filtering expectation $\mathbb{E}\left[H_t \mid Y_{1: t}=y_{1: t}\right]$ (over 50 runs and $N=100$). We considered the "true values" as the estimates of the filtering expectation obtained from a bootstrap PF with $N=10^{6}$.}
      \label{fig:inla1}
\end{figure}

\begin{flushleft}
where $t \in 1:T$ and $y_{t}$ is an observation of $Y_{t}$. We simulate data from this model using $T = 100$, $\sigma = 0.5$, $\alpha = 1$ and $\rho=0.7$. To fit the model with R-INLA \footnote{www.r-inla.org}, we considered these prior distributions for the parameters $\rho, \alpha, \sigma$, which are assumed to be unknown: $\tilde{\rho} \sim \mathcal{N} \left(m_{\rho},s_{\rho}^{2} \right), \alpha \sim \mathcal{N}(m_{\alpha},s_{\alpha}^{2}) $ and $\frac{1}{\sigma^{2}} \sim Gamma(a,b)$, where $\tilde{\rho} = log(1+\rho)-log(1-\rho)$ and we set a = b = 0.01, $m_{\rho}$ = 0, $s_{\rho}$ = 0.15, $m_{\alpha}$ = 0 and $s_{\alpha} = 10$. \\ Based on figures \ref{fig:3.01} and \ref{fig:3.002}, the particle filter that uses the proposal distribution constructed using INLA, to be referred to as INLA-based PF, yields log-likelihood estimates with smaller variability, when compared to the bootstrap PF. Interestingly, the variance of 50 log-likelihood estimates from the INLA-based PF, for $T=100$ and the simulated data from this model, is roughly the same as the one related to the bootstrap PF when using 1000 particles. Also for $T=500$, the INLA-based PF outperformed the bootstrap PF, regardless of the number of particles used where it is more apparent for $N=100$.\\

Figure \ref{fig:inla1} a) shows average Effective Sample Size (ESS) over time $t$ (ranging from 0 to 100) for both the INLA-based particle filter (in blue) and the bootstrap particle filter (in red), with 100 particles and 50 runs. The INLA-based particle filter appears to maintain relatively high ESS values over time, fluctuating around a higher range compared to the bootstrap particle filter, suggesting that the INLA-based filter may have better sample efficiency, especially when the ESS is closer to 100. On the other hand, the bootstrap particle filter shows more frequent and significant drops in ESS, sometimes dipping to values as low as 25 or lower. This implies that the bootstrap particle filter may suffer from degeneracy more often, where a few particles carry most of the weight, reducing its effective sample size. Although both particle filters exhibit fluctuations, the INLA-based particle filter appears to maintain a more stable and higher ESS on average.
\\
Another way to compare the performance of particle filters is to assess how they approximate the filtering expectation $\mathbb{E}\left[H_t \mid Y_{1: t}=y_{1: t}\right]$ can be approximated as follows
    \end{flushleft}

\begin{equation*}
\mathbb{E}\left[H_t \mid Y_{1: t}=y_{1: t}\right] \approx \sum_{i=1}^N W_{t}^i  H_t^i.
\end{equation*}
\begin{flushleft} 
 As we do not have the true values of $\mathbb{E}\left[H_t \mid Y_{1: t}=y_{1: t}\right]$ for all $t \in 1:T$ related to this model, we regarded the estimates of the filtering expectation obtained from a bootstrap particle filter with a very large number of particles, precisely $N=10^{6}$, as the true ones then we run the particle filters 50 times using 100 particles to estimate the filtering expectation at each time step $t \leq T$. Let $\Gamma_{t,j}^{N}$ denote the estimate of $\mathbb{E}\left[H_t \mid Y_{1: t}=y_{1: t}\right]$ using $N$ particles in the $j$-th run of a particle filter. We considered the average absolute difference $\Sigma_{j=1}^{50} \frac{1}{50}|\Gamma_{t,j}^{10^{2}} - \Gamma_{t,j}^{10^{6}}|$ for each time step $t$ as a diagnostic metric. \\

Based on figure \ref{fig:inla1} b), both lines show fluctuations over time but the blue line representing the INLA-based particle filter consistently shows lower average absolute errors compared to the bootstrap particle filter throughout the time range. This suggests that the INLA-based method provides more accurate filtering estimates, with smaller deviations from the "true values" on average.

\end{flushleft} 


\newpage
\subsection{Particle Marginal Metropolis Hastings}
\subsubsection{Initialisation of Particle Marginal Metropolis Hastings using INLA}

\begin{flushleft}
We simulated data from the previous model (\ref{model}) using $T = 100$, $\sigma = 0.5$, $\alpha = 0.5$ and $\rho=0.85$, we implemented PMMH based on a proposal that is a Gaussian random walk, with covariance $\epsilon^{2}\mathbf{I}$ where $\mathbf{I}$ is a $3 \times 3$ identity matrix and the bootstrap particle filter associated to the (\ref{model}) model, where $N = 100$ and $\epsilon =$ 0.3. The number of iterations is set to $10^4$ and the burn-in period to $10^3$. We also thin the chain by taking every 10-th observation. \\
Initializing the Markov chain(s) in areas of high posterior mass is important, in the case of PMMH samplers \parencite{pmmh1}, which might result in faster exploration of those regions. We believe that using INLA, whenever possible, is a good practice in this context so we set initial values, $\theta_{init}$, to the posterior modes of the corresponding parameters, estimated by INLA. 
As shown in Table \ref{tab:1}, the posterior modes obtained from the PMMH sampler (where the bootstrap particle filter was used to estimate the marginal likelihood) are closer to the true values used for simulating the data when compared to the ones obtained from a single run of INLA, especially when it comes to the standard deviation $\sigma$. \\
\end{flushleft}

\begin{table}[h]
\centering
\begin{tabular}{lllll}
\cline{2-4}
\multicolumn{1}{l|}{}  & \multicolumn{1}{l|}{$\rho$} & \multicolumn{1}{l|}{$\alpha$} & \multicolumn{1}{l|}{$\sigma$} &  \\ \cline{1-4}
\multicolumn{1}{|l|}{INLA} & \multicolumn{1}{l|}{0.80 (0.57,0.96)} & \multicolumn{1}{l|}{0.58 (-0.1,1.5)} & \multicolumn{1}{l|}{0.91(0.47,1.32)} &  \\ \cline{1-4}
\multicolumn{1}{|l|}{PMH-BPF} & \multicolumn{1}{l|}{0.85 (0.77,0.94)} & \multicolumn{1}{l|}{0.5 (0.14,0.76) } & \multicolumn{1}{l|}{0.59(0.44,0.71)} &  \\ \cline{1-4}
                       &                       &                       &                       & 
\end{tabular}
\caption{Modes (and 95\% credible intervals) of the marginal posterior distributions of the parameters obtained using INLA (first row) and PMMH sampler(second row) based a Bootstrap particle filter with $N=100$ and $\epsilon = 0.3$.}
\label{tab:1} 
\end{table}


\subsubsection{Using the particle filter with the suggested proposal distributions in Particle Marginal Metropolis Hastings}

\begin{flushleft}
    We would like to learn the parameters $\theta = (\rho, \alpha, \sigma)$ of the model (\ref{model}) using the PMMH sampler. We considered two cases where the marginal likelihood was estimated by the INLA-based PF and the bootstrap particle filter (BPF). We used the same set-up for data simulation as in the previous experiment. Each chain consisted of 10000 iterations. Similar to the previous experiment, the burn-in period and the thinning parameter were set to $10^3$ and to 10 respectively. To compare the performance of the BPF and INLA-based PF in the context of particle MCMC, we took into consideration the acceptance probability and the the estimates of integrated autocorrelation time (IAT). The latter is often used as a diagnostic metric for MCMC samplers. It measures the number of iterations or time steps required for a Markov chain to become approximately independent (IAT=1 if all MCMC samples are independent). In other words, it quantifies the rate at which successive samples in a chain become uncorrelated with each other. Minimizing the integrated autocorrelation time is of paramount importance in MCMC-based algorithms as it directly impacts the effectiveness of these methods.
    
    We picked different values of $\epsilon=\{0.05,0.1,0.25,1.25\}$. The task of tuning is undoubtedly time-consuming and there is no definite way of doing it correctly. In case of setting $\epsilon$ to a very small value, it would lead to a high correlation between the samples and extremely slow mixing of the chain. On the other hand, choosing a relatively high value would lead to long jumps in the posterior region which can result in low acceptance probabilities and poor mixing of the chain. However, setting $\epsilon$ to a comparatively low value and starting the chain from the posterior mode of the parameter or from a value that is close to it (which is similar to what we did in the previous experiment using the posterior summaries obtained from INLA) can potentially lead to an adequate performance as illustrated in Table \ref{tab:2} which presents the acceptance rates and integrated autocorrelation time  (IAT) estimates of the parameters’ chains
for PMMH based on the bootstrap particle filter and the INLA-based Particle filter

    Table \ref{tab:2} shows that using the particle filter based on the proposal distribution approximated by INLA, within PMMH leads to an improvement, for suitable tuning parameters, with respect to the PMMH which is based on the bootstrap particle filter as the latter resulted in acceptance probabilities that are lower than the ones related to PMMH with INLA-based PF. In addition, the IATs of the parameters' chains of the PMMH with a bootstrap particle filter, are generally higher than those of the PMMH with the INLA-based particle filter. Choosing a supposedly large tuning parameter ($\epsilon=1.25$) led to a low performance regardless of the proposal of the particle filter used. 
\end{flushleft}


\begin{table}[h]
\centering
\begin{tabular}{|ll|ll|ll|ll|ll|}
\hline
                       & $\epsilon$ &                       & 0.05 &                      & 0.1  &                     & 0.25 &                       & 1.25 \\ \hline
                       & Proposal of the particle filter& \multicolumn{1}{l|}{INLA} & BPF & \multicolumn{1}{l|}{INLA} &  BPF & \multicolumn{1}{l|}{INLA} & BPF & \multicolumn{1}{l|}{INLA} & BPF \\ \hline
                       & Acceptance rate & \multicolumn{1}{l|}{0.643} & 0.425 & \multicolumn{1}{l|}{0.527} & 0.362 & \multicolumn{1}{l|}{0.281} & 0.202 & \multicolumn{1}{l|}{0.06} & 0.07 \\ \hline
\multicolumn{1}{|l|}{} & $\rho$ & \multicolumn{1}{l|}{1.082} & 1.825  & \multicolumn{1}{l|}{1.211} & 1.921 & \multicolumn{1}{l|}{2.101} & 2.236 & \multicolumn{1}{l|}{4.782} & 3.838 \\ \cline{2-10} 
\multicolumn{1}{|l|}{IAT} & $\sigma$ & \multicolumn{1}{l|}{1.134} & 1.64 & \multicolumn{1}{l|}{1.196} & 1.841 & \multicolumn{1}{l|}{2.149} & 2.45 & \multicolumn{1}{l|}{4.682} & 4.332 \\ \cline{2-10} 
\multicolumn{1}{|l|}{} & $\alpha$ & \multicolumn{1}{l|}{1.029} & 1.78 & \multicolumn{1}{l|}{1.116} & 1.944 & \multicolumn{1}{l|}{2.029} & 2.56 & \multicolumn{1}{l|}{2.173} & 2.82 \\ \hline
\end{tabular}
\caption{Acceptance rates  and estimates of integrated autocorrelation time (IAT) of the parameters' chains for PMMH  based on the bootstrap particle filter and the INLA-based Particle filter (where $N=200$), for
different values of $\epsilon$=\{0.05,0.1,0.25,1.25\}. The time horizon was $T=100$, the number of iterations used were $10^4$ (burn-in iterations were $10^3$) and thinning was set to 10.}
\label{tab:2}

\end{table}
   
\begin{table}[]
\centering
\begin{tabular}{|ll|l|l|}
\hline
Proposal of the particle filter             &  & INLA & BPF\\ \hline
Acceptance rate                       &  & 0.173 &  0.135\\ \hline
\multicolumn{1}{|l|}{} & $\rho$  & 2.02 & 2.981 \\ \cline{2-4} 
\multicolumn{1}{|l|}{IAT} &  $\sigma$ &  2.15& 2.832 \\ \cline{2-4} 
\multicolumn{1}{|l|}{} & $\alpha$ & 1.85 & 3.101 \\ \hline
\end{tabular}
\caption{Acceptance rates and estimates of integrated autocorrelation time (IAT) of parameter chains for PMMH, using the bootstrap particle filter ($N=400$) and INLA-based particle filter ($N=100$), with $\epsilon$=0.1, time horizon $T=400$, 10,000 iterations (1,000 burn-in, thinning of 10).}

\label{tab:3} 

\end{table}


\begin{flushleft}
    
\end{flushleft}

\begin{figure}
  \centering
  \begin{subfigure}{0.33\linewidth}
    \centering
    \includegraphics[height = 3.25cm,width=\linewidth]{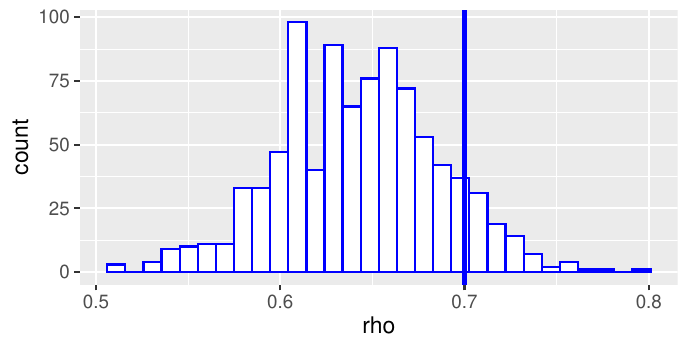}
    \caption{$\rho$}
  \end{subfigure}%
  \begin{subfigure}{0.33\linewidth}
    \centering
    \includegraphics[height = 3.25cm,width=\linewidth]{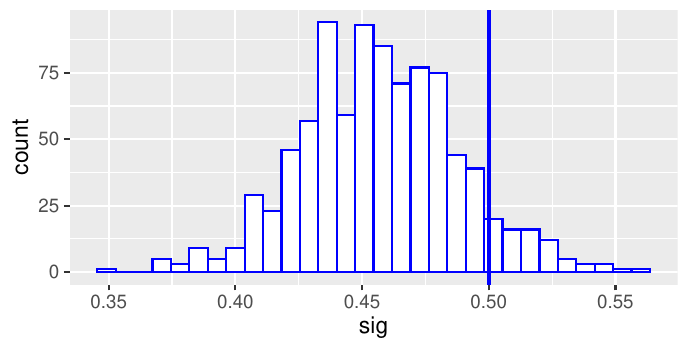}
    \caption{$\sigma$}
  \end{subfigure}%
  \begin{subfigure}{0.33\linewidth}
    \centering
    \includegraphics[height = 3.25cm,width=\linewidth]{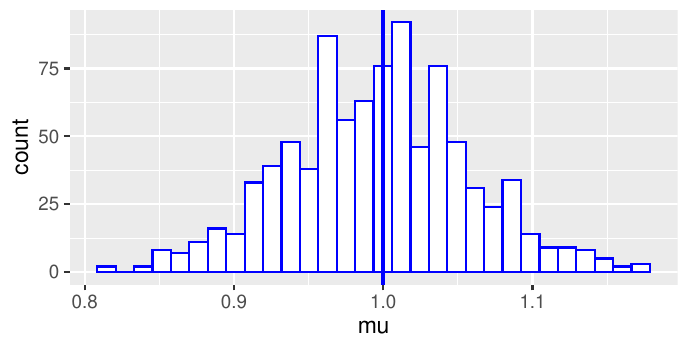}
    \caption{$\alpha$}
  \end{subfigure}

  \begin{subfigure}{0.33\linewidth}
    \centering
    \includegraphics[height = 3.3cm,width=\linewidth]{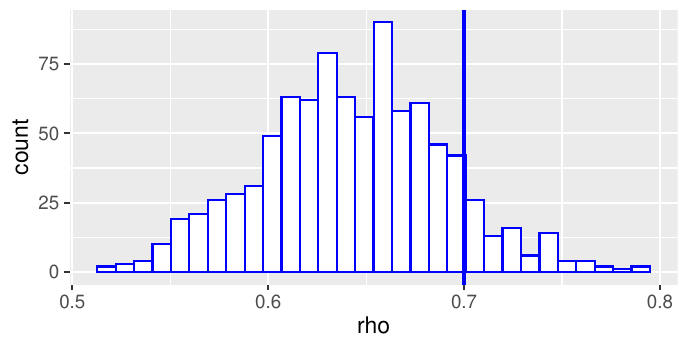}
    \caption{$\rho$}
  \end{subfigure}%
  \begin{subfigure}{0.33\linewidth}
    \centering
    \includegraphics[height = 3.3cm,width=\linewidth]{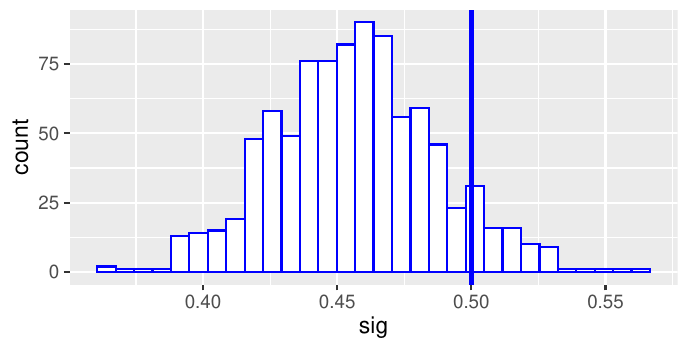}
    \caption{$\sigma$}
  \end{subfigure}%
  \begin{subfigure}{0.33\linewidth}
    \centering
    \includegraphics[height = 3.3cm,width=\linewidth]{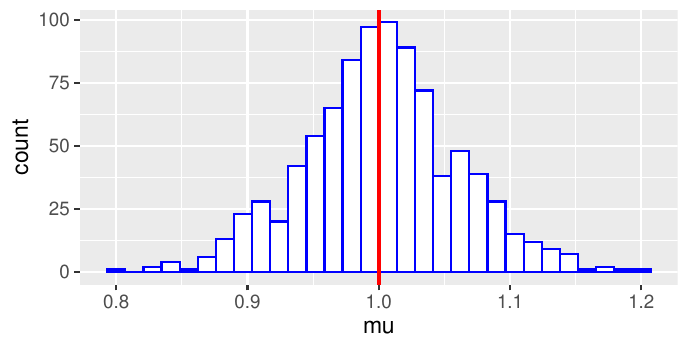}
    \caption{$\alpha$}
  \end{subfigure}

  \caption{Marginal posterior distributions of the parameters, The first row is related to the PMMH sampler where the bootstrap particle filter was used while the second row is related to he PMMH sampler where the INLA-based PF was used. Vertical dark lines represent the true values used
for simulating the data set.}
  \label{fig:subfigures}
\end{figure}

\begin{flushleft}
In a different experiment, we extended our analysis to a longer time horizon, $T$. It is important to note that the variance of the estimates for the marginal likelihood increases as $T$ grows, which implies that a higher number of particles is required to achieve satisfactory performance in the PMMH sampler. Specifically, for the BPF, we set the number of particles, $N$, equal to $T$, while for the INLA-based PF, we limited it to 100 particles. The results of this experiment are summarized in both Table \ref{tab:3} and Figure \ref{fig:subfigures}, revealing an increase in efficiency when using the INLA-based PF in the PMMH sampler as compared to the BPF, with just one-fourth of the total number of particles used in the latter. Needless to mention, initiating the Markov chains from locations distant from the posterior modes will result in higher IAT, especially when the tuning process is not done properly.
\end{flushleft}

\section{Conclusions}
\label{conclusions}
\begin{flushleft}
    We proposed a way to build a proposal distribution of the particle filter based on an approximation using integrated nested Laplace approximation (INLA). We have demonstrated that this method performed well on simulated data from a Poisson state-space model with Gaussian transition distribution. 
The main three contributions of the paper are:
  1) \emph{Improving the proposal distribution in a particle filter}: The main two metrics used to evaluate the performance of the proposal distributions were the variance of the marginal likelihood estimates and the effective sample size (ESS). Considering the previously mentioned indicators, the suggested method demonstrated superior performance compared to the bootstrap particle filter, particularly when dealing with longer time horizons (T) and a significantly reduced number of particles.
  2) \emph{Parameter estimation}: incorporating the particle filter with INLA-based proposal within the PMMH sampler may potentially enhance the mixing of the chains as the variance of the marginal likelihood estimates is reduced compared to those obtained using a bootstrap particle filter. On another note, INLA can serve as an "offline" approach for estimating parameters in particular state-space models in addition to the techniques outlined in \parencite{conc1,kantas2009overview,conc2,johansen2015blocks,johansen2008particle, johansen2020sequential}.
3) \emph{Initialization of the Markov chains}: As shown, INLA can also be combined with Particle MCMC methods (in our case, we considered the PMMH sampler) which has two major advantages: a) PMMH can potentially improve the approximation of INLA. b) less chains' iterations would be required when using INLA for initializing the chains.

However, it is worth reiterating that the suggested approach can work only if the state-space model is a case of a latent Gaussian model. This specific class of models is prevalent in the literature, both in terms of its applications and the instances used for simulation. Nevertheless, state-space models where the transition distribution is not Gaussian, which are also of interest in many cases, can not be considered in our framework.

 Furthermore, another limitation of our work consists of the fact that the proposed method is computationally more expensive than the bootstrap particle filter mainly due to the calculations in (\ref{eq4.3}) as they involve calculating the inverse of the covariance matrix of the size $(t-1) \times (t-1)$ for all $t \in [3,T]$  (assuming that $3 \leq T$), there is also a memory cost related to the storage of the precision matrices. In addition, the use of INLA might not lead to an improvement in case of having a high-dimensional parameter space and/or parameters with multi-modal posterior distributions. For future work, it would be valuable to compare this approach with other types of particle filters and proposal distributions. Further, designing proposal distributions based on INLA within a conditional particle filter framework and applying this in particle Gibbs would be an intriguing direction to explore.
\end{flushleft}

\section*{Acknowledgements}
\begin{flushleft}
    This work was motivated by discussions with Nicolas Chopin. The author would also like to thank Finn Lindgren, Víctor Elvira, Amy Wilson and Håvard Rue for very helpful comments. The author is supported by Edinburgh Future Cities studentship.
\end{flushleft}

\printbibliography 

@article{johansen2020sequential,
  author = {Johansen, A. M.},
  title = {Sequential Monte Carlo: Particle Filtering and Beyond},
  year = {2020}
}

@article{kantas2009overview,
  author = {Kantas, N. and Doucet, A. and Singh, S. S. and Maciejowski, J. M.},
  title = {An Overview of Sequential Monte Carlo Methods for Parameter Estimation in General State-Space Models},
  journal = {IFAC Proceedings Volumes},
  volume = {42},
  number = {10},
  pages = {774--785},
  year = {2009}
}

@article{johansen2012exact,
  author = {Johansen, A. M. and Whiteley, N. and Doucet, A.},
  title = {Exact Approximation of Rao-Blackwellised Particle Filters},
  journal = {IFAC Proceedings Volumes},
  volume = {45},
  number = {16},
  pages = {488--493},
  year = {2012}
}

@incollection{whiteley2010recent,
  author = {Whiteley, N. and Johansen, A. M.},
  title = {Recent Developments in Auxiliary Particle Filtering},
  booktitle = {Inference and Learning in Dynamic Models},
  pages = {39--47},
  volume = {38},
  year = {2010}
}

@article{johansen2008particle,
  author = {Johansen, A. M. and Doucet, A. and Davy, M.},
  title = {Particle Methods for Maximum Likelihood Estimation in Latent Variable Models},
  journal = {Statistics and Computing},
  volume = {18},
  pages = {47--57},
  year = {2008}
}

@article{johansen2015blocks,
  author = {Johansen, A. M.},
  title = {On Blocks, Tempering and Particle MCMC for Systems Identification},
  journal = {IFAC-PapersOnLine},
  volume = {48},
  number = {28},
  pages = {969--974},
  year = {2015}
}

@article{johansen2008note,
  author = {Johansen, A. M. and Doucet, A.},
  title = {A Note on Auxiliary Particle Filters},
  journal = {Statistics \& Probability Letters},
  volume = {78},
  number = {12},
  pages = {1498--1504},
  year = {2008}
}

@article{bpf,
  author = {N. Gordon and D. Salmond and A. Smith},
  title = {Novel approach to nonlinear/non-Gaussian Bayesian state estimation},
  journal = {IEE Proceedings F (Radar And Signal Processing)},
  volume = {140},
  year = {1993}
}

@article{rbpf,
  author = {A. Doucet and S. Godsill and C. Andrieu},
  title = {In Statistics and Computing},
  journal = {Springer Science and Business Media LLC},
  volume = {10},
  number = {3},
  pages = {197--208},
  year = {2000}}

@article{apf,
  author = {Pitt, M. and Shephard, N.},
  title = {Filtering via Simulation: Auxiliary Particle Filters},
  journal = {Journal of the American Statistical Association},
  volume = {94},
  number = {446},
  pages = {590--599},
  year = {1999}
}

@article{apf2,
  author = {P. Guarniero and A. Johansen and A. Lee},
  title = {The Iterated Auxiliary Particle Filter},
  journal = {Journal of the American Statistical Association},
  volume = {112},
  pages = {1636--1647},
  year = {2017}
}

@inproceedings{apf5,
  author = {C. Andrieu and M. Davy and A. Doucet},
  title = {Improved Auxiliary Particle Filtering: Applications To Time-Varying Spectral Analysis},
  booktitle = {Proceedings of the 11th IEEE Signal Processing Workshop on Statistical Signal Processing},
  year = {2001}
}

@inproceedings{apf3,
  author = {V. Elvira and L. Martino and M. Bugallo and P. Djuri{\'c}},
  title = {In Search for Improved Auxiliary Particle Filters},
  booktitle = {2018 26th European Signal Processing Conference (EUSIPCO)},
  pages = {1637--1641},
  year = {2018}
}

@inproceedings{apf4,
  author = {N. Branchini and V. Elvira},
  title = {Optimized auxiliary particle filters: adapting mixture proposals via convex optimization},
  booktitle = {Proceedings of the Thirty-Seventh Conference on Uncertainty in Artificial Intelligence},
  volume = {161},
  pages = {1289--1299},
  year = {2021},
  month = {7},
  day = {27}
}

@book{int3,
  author = {A. Doucet and J. F. G. de Freitas and N. J. Gordon},
  title = {Sequential Monte Carlo Methods in Practice},
  publisher = {Springer-Verlag},
  address = {New York},
  year = {2001}
}

@article{int0,
  author = {N. Shephard and M. Pitt},
  title = {Likelihood analysis of non-Gaussian measurement time series},
  journal = {Biometrika},
  volume = {84},
  number = {3},
  pages = {653--667},
  year = {1997}
}

@book{int1,
  author = {O. Capp{\'e} and E. Moulines and T. Ryd{\'e}n},
  title = {Inference in Hidden Markov Models},
  publisher = {Springer},
  year = {2005}
}

@article{int5,
  author = {A. Doucet and S. Godsill and C. Andrieu},
  title = {On sequential Monte Carlo sampling methods for Bayesian filtering},
  journal = {Statistics and Computing},
  volume = {10},
  pages = {197--208},
  year = {2000}
}

@inproceedings{int4,
  author = {R. Merwe and A. Doucet and N. Freitas and E. Wan},
  title = {The Unscented Particle Filter},
  booktitle = {Advances in Neural Information Processing Systems},
  volume = {13},
  year = {2000}
}

@book{Chopin2020,
  author = {N. Chopin and O. Papaspiliopoulos},
  title = {An Introduction to Sequential Monte Carlo},
  publisher = {Springer International Publishing},
  year = {2020}}

@article{inla1,
  author = {H. Rue and S. Martino and N. Chopin},
  title = {Approximate Bayesian inference for latent Gaussian models by using integrated nested Laplace approximations},
  journal = {Journal of the Royal Statistical Society: Series B (Statistical Methodology)},
  volume = {71},
  pages = {319--392},
  year = {2009}
}

@article{inla01,
  author = {H. Rue and A. Riebler and S. H. S{\o}rbye and J. B. Illian and D. Simpson and F. Lindgren},
  title = {Bayesian Computing with INLA: A Review},
  journal = {Annual Review of Statistics and Its Application},
  volume = {4},
  number = {1},
  pages = {395--421},
  year = {2017}}

@article{inla02,
  author = {T. G. Martins and D. Simpson and F. Lindgren and H. Rue},
  title = {Bayesian computing with INLA: New features},
  journal = {Computational Statistics Data Analysis},
  volume = {67},
  pages = {68--83},
  year = {2013}
}

@article{inla03,
  author = {F. Lindgren and H. Rue and J. Lindstr{\"o}m},
  title = {An explicit link between Gaussian fields and Gaussian Markov random fields: the stochastic partial differential equation approach},
  journal = {Journal of the Royal Statistical Society: Series B (Statistical Methodology)},
  volume = {73},
  pages = {423--498},
  year = {2011}
}

@article{inla04,
  author = {R. Ruiz-C{\'a}rdenas and E. Krainski and H. Rue},
  title = {Direct fitting of dynamic models using integrated nested Laplace approximations — INLA},
  journal = {Computational Statistics \& Data Analysis},
  volume = {56},
  pages = {1808--1828},
  year = {2012}
}

@book{inla05,
  author = {N. Ravishanker and B. Raman and R. Soyer},
  title = {Dynamic Time Series Models using R-INLA: An Applied Perspective},
  publisher = {Chapman and Hall/CRC},
  year = {2022}}

@book{inla2,
  author = {H. Rue and L. Held},
  title = {Gaussian Markov Random Fields: Theory and Applications},
  publisher = {Chapman \& Hall},
  year = {2005}}

@article{pmcmc,
  author = {C. Andrieu and A. Doucet and R. Holenstein},
  title = {Particle Markov chain Monte Carlo methods},
  journal = {Journal of the Royal Statistical Society: Series B (Statistical Methodology)},
  volume = {72},
  number = {3},
  pages = {269--342},
  year = {2010}
}

@article{rue2017bayesian,
  title={Bayesian computing with INLA: a review},
  author={Rue, H. and Riebler, A. and Sørbye, S. H. and Illian, J. B. and Simpson, D. P. and Lindgren, F. K.},
  journal={Annual Review of Statistics and Its Application},
  volume={4},
  number={1},
  pages={395--421},
  year={2017}
}

@article{azzalini1999statistical,
  title={Statistical Applications of the Multivariate Skew Normal Distribution},
  author={Azzalini, A. and Capitanio, A.},
  journal={Journal of the Royal Statistical Society: Series B (Statistical Methodology)},
  volume={61},
  number={3},
  pages={579--602},
  year={1999}}

@article{defreitas2000sequential,
  title={Sequential Monte Carlo methods to train neural network models},
  author={de Freitas, J. F. G. and Niranjan, M. and Gee, A. H. and Doucet, A.},
  journal={Neural Computation},
  volume={12},
  number={4},
  pages={955--993},
  year={2000}
}

@inproceedings{vandermerwe2000unscented,
  title={The unscented particle filter},
  author={Van Der Merwe, R. and Doucet, A. and De Freitas, N. and Wan, E.},
  booktitle={Advances in Neural Information Processing Systems},
  volume={13},
  year={2000}
}

@book{delmoral,
  author = {Del Moral, P.},
  title = {Feynman-Kac Formulae: Genealogical and Interacting Particle Systems with Applications},
  publisher = {Springer-Verlag},
  address = {New York},
  year = {2004}
}

@article{delmoral2,
  author = {F. C{\'e}rou and P. Del Moral and A. Guyader},
  title = {A non asymptotic variance theorem for unnormalized Feynman-Kac particle models},
  journal = {Annales de l'Institut Henri Poincar{\'e}},
  volume = {47},
  pages = {629--649},
  year = {2011}
}

@article{pmmh1,
  author = {J. Dahlin and T. B. Sch{\"o}n},
  title = {Getting Started with Particle Metropolis-Hastings for Inference in Nonlinear Dynamical Models},
  journal = {Journal of Statistical Software},
  volume = {88},
  number = {2},
  pages = {1--41},
  year = {2019}}

@article{conc1,
  author = {N. Kantas and A. Doucet and S. Singh and J. Maciejowski and N. Chopin},
  title = {On Particle Methods for Parameter Estimation in State-Space Models},
  journal = {Statistical Science},
  volume = {30},
  pages = {328--351},
  year = {2015}
}

@article{conc2,
  author    = {D. Luengo and L. Martino and M. Bugallo and V. Elvira and S. S{\"a}rkk{\"a}},
  title     = {A survey of Monte Carlo methods for parameter estimation},
  journal   = {EURASIP Journal on Advances in Signal Processing},
  year      = {2020},
  volume    = {2020},
  number    = {25}}
\end{document}